\documentclass[pra,twocolumn,superscriptaddress]{revtex4-1}

\usepackage{amsmath}    
\usepackage{graphicx}   
\usepackage{hhline}
\usepackage{verbatim}   
\usepackage{color}      
\usepackage{hyperref}   
\usepackage{longtable}
\usepackage[caption=false]{subfig}
\usepackage{grffile}
\usepackage{changepage}
\usepackage{siunitx}
\newcommand\numberthis{\addtocounter{equation}{1}\tag{\theequation}}

\begin{document}

\title{Photoassociation of Ultracold NaLi}

\author{Timur M. Rvachov}
\affiliation{Research Laboratory of Electronics, MIT-Harvard Center for Ultracold Atoms,
Department of Physics, Massachusetts Institute of Technology, Cambridge, Massachusetts 02139, USA}

\author{Hyungmok Son}
\affiliation{Research Laboratory of Electronics, MIT-Harvard Center for Ultracold Atoms,
Department of Physics, Massachusetts Institute of Technology, Cambridge, Massachusetts 02139, USA}
\affiliation{Department of Physics, Harvard University, Cambridge, Massachusetts 02138, USA}

\author{Juliana J. Park}
\affiliation{Research Laboratory of Electronics, MIT-Harvard Center for Ultracold Atoms,
Department of Physics, Massachusetts Institute of Technology, Cambridge, Massachusetts 02139, USA}

\author{Pascal M. Notz}

\affiliation{Research Laboratory of Electronics, MIT-Harvard Center for Ultracold Atoms,
Department of Physics, Massachusetts Institute of Technology, Cambridge, Massachusetts 02139, USA}
\affiliation{Institut f{\"u}r Angewandte Physik, Technische Universit{\"a}t Darmstadt, Darmstadt, Germany}

\author{Tout T. Wang}
\affiliation{Research Laboratory of Electronics, MIT-Harvard Center for Ultracold Atoms,
Department of Physics, Massachusetts Institute of Technology, Cambridge, Massachusetts 02139, USA}
\affiliation{Department of Physics, Harvard University, Cambridge, Massachusetts 02138, USA}

\author{Martin W. Zwierlein}
\affiliation{Research Laboratory of Electronics, MIT-Harvard Center for Ultracold Atoms,
Department of Physics, Massachusetts Institute of Technology, Cambridge, Massachusetts 02139, USA}

\author{Wolfgang Ketterle}
\affiliation{Research Laboratory of Electronics, MIT-Harvard Center for Ultracold Atoms,
Department of Physics, Massachusetts Institute of Technology, Cambridge, Massachusetts 02139, USA}

\author{Alan O. Jamison}
\affiliation{Research Laboratory of Electronics, MIT-Harvard Center for Ultracold Atoms,
Department of Physics, Massachusetts Institute of Technology, Cambridge, Massachusetts 02139, USA}

\date{\today}

\begin{abstract}

We perform photoassociation spectroscopy in an ultracold $^{23}$Na-$^6$Li mixture to study the $c^3\Sigma^+$ excited triplet molecular potential. We observe 50 vibrational states and their substructure to an accuracy of 20 MHz, and provide line strength data from photoassociation loss measurements. An analysis of the vibrational line positions using near-dissociation expansions and a full potential fit is presented.  This is the first observation of the $c^3\Sigma^+$ potential, as well as photoassociation in the NaLi system.

\end{abstract}

\maketitle

\section{Introduction}

The process of photoassociation (PA) transfers two unbound atoms into a molecular state.  It connects systems of ultracold atoms (typically colder than 1\,\,$\mu$K), which can be prepared and controlled with the methods of atomic physics, to the molecular Hilbert space. This connection can be used to manipulate the atomic systems themselves, or explore molecular physics from an initial atomic state.  In atoms, photoassociation resonances can tune atomic interactions via optical Feshbach resonances \cite{optical_fesh_review}, and photoassociative loss can modify many-body dynamics via the quantum Zeno effect \cite{pa_takahashi}. From a molecular perspective, photoassociation has seen widespread use in production of cold dipolar molecules \cite{pa_formation_lics, pa_formation_nacs, pa_formation_krb, triplet_rbcs_dulieu, pa_formations_rbcs, inouye_pa_stirap} and molecular spectroscopy, where the very small kinetic energy spread and precise state control in ultracold systems can provide an advantage over traditional spectroscopy techniques \cite{pa_julienne,stwalley_paspec}. Such spectroscopic knowledge has been applied in production of near-quantum degenerate dipolar molecules \cite{krb_ni_2008,rbcs_hcn_2014,rbcs_cornish_2014, nak_park_2015, narb_wang_2016, us_groundstate}, which open the door to studies of molecular dynamics and chemistry in the quantum regime \cite{chemistry_review_krems_2008,review_ye_2009}.

Photoassociation has been studied for many heteronuclear combinations of atomic species that have been laser cooled \cite{heteronuclear_pa_weidemuller}. One of the few unstudied exceptions is the case of $^{23}$Na and $^6$Li, which are among the workhorse species in the field of ultracold atoms. $^{23}$Na$^6$Li forms the lightest and in many regards the simplest dipolar bi-alkali molecule. Triplet excited states of the NaLi molecule have never been spectroscopically observed. Searches for such transitions in hot thermal ensembles were not successful since such ensembles contain predominantly singlet molecules, and the small spin-orbit coupling in NaLi makes triplet-singlet mixing among excited molecular potentials very weak \cite{nali_Xsigma_2}.
 
In this work we perform photoassociation of NaLi and map out 50 vibrational states in the $c^3\Sigma^+$ excited molecular triplet potential. Our entry point into the triplet manifold is free Na and Li atoms prepared in atomic states with total electron spin 1.  Our motivation was to obtain spectroscopic information necessary to identify a pathway for creating ultracold NaLi molecules in the absolute triplet ground state. Using the results reported in this paper and the following report \cite{ourtwophoton}, we have recently reached this goal \cite{us_groundstate}.  Triplet NaLi molecules are an attractive system for exploring new regimes of many-body physics due to a unique combination of properties:  They have a predicted electric dipole moment of 0.2\,D \cite{dmoment_krems_2013}, a magnetic dipole moment of $2\,\mu_B$, and  a small inelastic collision rate, allowing for long lifetimes even in the metastable triplet state \cite{univrates_julienne_2011,us_groundstate}.

\section{Photoassociation spectroscopy}
We produce a dual species mixture of Na and Li in the $|F\,{=}\,2,m_F\,{=}\,2\rangle$ and $|F\,{=}\,3/2,m_F\,{=}\,3/2\rangle$ angular momentum states, respectively,  confined in a cigar shaped Ioffe-Pritchard magnetic trap with trapping frequencies for Na of $\omega_r \,{=}\, 2\pi \,{\times}\, 325\,$Hz, $\omega_z \,{=}\,2\pi\,{\times}\, 25\,$Hz and the corresponding trap frequencies for Li scaled by the mass ratio $\sqrt{m_\text{Na}/m_\text{Li}}\,{\approx}\, 2$. The sample contains $N_\text{Na} \,{=}\, 1 \,{\times}\, 10^7$ sodium atoms and $N_\text{Li} \,{=}\, 4 \,{\times}\, 10^6$ lithium atoms at a temperature of $T\,{=}\,3.7\,\mu\text{K}$. This corresponds to peak densities of $n_\text{Na}\,{=}\,9 \,{\times}\, 10^{12}\,\text{cm}^{-3}$ and $n_\text{Li}\,{=}\,3 \,{\times}\, 10^{12}\,\text{cm}^{-3}$, with degeneracy parameters $T/T_c\,{=}\,2.7$ and $T/T_F\,{=}\,1$, where $T_c$ is the Na condensation temperature and $T_F$ is the Li Fermi temperature.  The bias field of the magnetic trap is 1.2\,G aligned along the long ($z$) axis of the trap. The free atoms are detected by imaging the absorption of resonant light. The angular momentum state of a colliding Na-Li pair is $|J=1, m_J=1\rangle$, where $J\,{=}\,N+S$ is the total angular momentum excluding nuclear spin, $S\,{=}\,1$ is the total electronic spin of the two-body system, and $N\,{=}\,0$ is the orbital angular momentum of $s$-wave collisions, which is the dominant collisional partial wave at our temperatures.

\begin{figure}[!ht]
\includegraphics[width=0.5\textwidth]{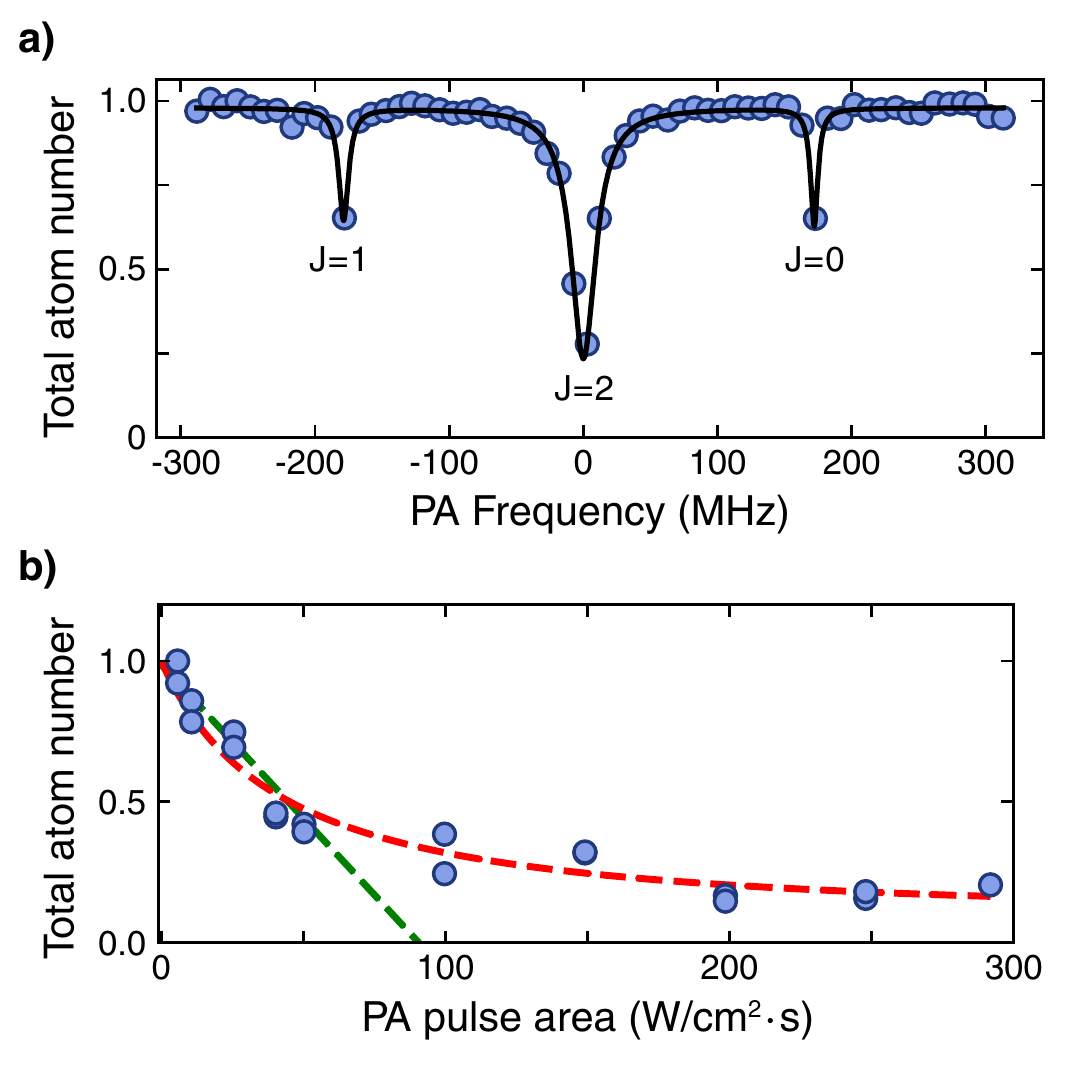}
\caption{Sample PA data for the $v\,{=}\,11,N\,{=}\,1$ state of the $c^3\Sigma^+$ potential. a) The loss spectrum in the total ($N_\text{Na}+N_\text{Li}$) atom number as the PA frequency is varied, showing a multiplet line structure.  b) The remaining total atom number with the PA beam tuned to the strongest (central) line, as a function of PA pulse area. The data is fit to a linear (green) and two-body (red) loss model, from which we extract the PA line strengths.}
\label{fig:sampleline}
\end{figure}

The ultracold mixture is illuminated with a PA beam along the bias field ($z$) direction with linear polarization (typical beam waist of $250\,\mu\text{m}$, power up to $300\,$mW). The PA light is produced by one of two tunable Ti:Sapphire lasers, covering the wavelength range $670\,{-}\,725\,$nm and  $725\,{-}\,1000\,$nm. If the PA beam is tuned on resonance with an electronically excited molecular state, the atoms are photoassociated to this excited state and then spontaneously decay down into molecular bound states. We detect only the remaining free atoms, hence PA resonance is observed by simultaneous loss in both Na and Li free atoms. The PA exposure was 10 to 4000\,ms, long enough to see atom loss depending on the strength of the PA line.  During exposure the PA laser is swept over a range of 10-500\,\,MHz depending on the fineness of the spectroscopic search. 

We fit the loss spectra in total atom number, $N_\text{Na}+N_\text{Li}$, to extract the PA line center (Fig.\,\ref{fig:sampleline}a).
\begin{figure}[!ht]
\includegraphics[width=0.5\textwidth]{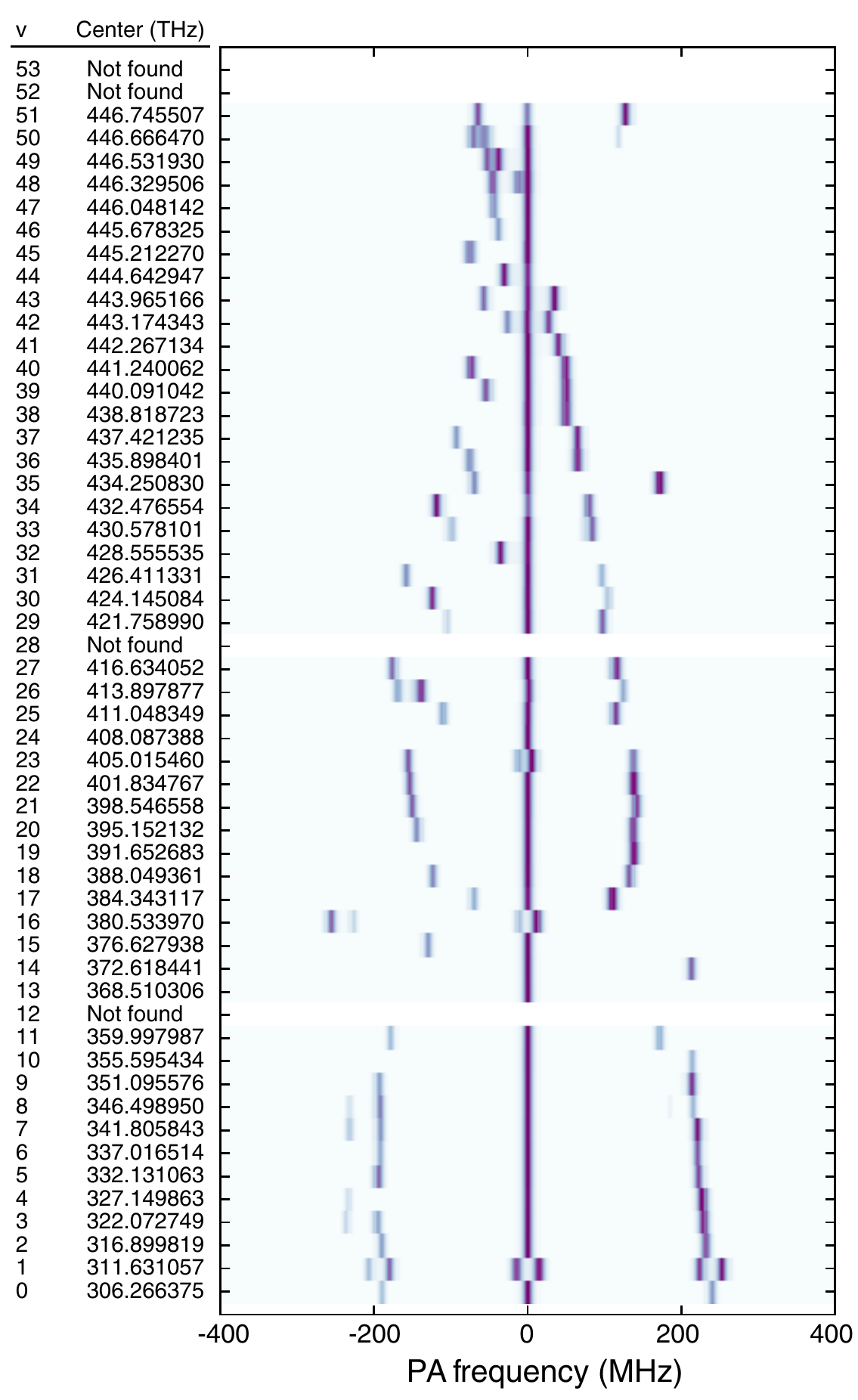}
\caption{Fitted PA line positions and observed substructure. For each vibrational level, multiple finely split lines were observed. The center of the line positions are reported on the left, which correspond to the center (zero detuning) lines in the substructure diagram on the right. Binding energies can be calculated by taking the difference of the PA laser frequency and the Li D1 asymptote, 446.789528\,THz \cite{porto_li_spec}. The choice of line center was made by visual inspection, by considering the overall symmetry of the manifold as well as any obvious doublets. The strength data within a vibrational level was insufficient to define a meaningful spectroscopic center of mass. Aside from the $v\,{=}\,11$ line, we have not made a careful identification of the quantum numbers associated with the substructure, and present the data raw. However, a clear increase in substructure splittings is observed towards lower $v$, consistent with short-range interaction effects which may intensify in deeply bound molecular states. Note that this figure shows the fitted Lorentzians with the linewidth artificially set to 9\,\,MHz: This is done for clarity as PA spectra with large loss are broadened by depletion of the sample density.}\label{fig:allspec}
\end{figure}
The PA loss feature is observed in both Na and Li atom numbers, however fitting the total atom number improves the signal-to-noise of the PA feature because of technical anti-correlated number fluctuations in the experiment arising from sympathetic cooling \cite{zoran_fermions}. The loss features were fit to Lorentzian lineshapes, and by fitting spectra with low PA loss (avoiding density depletion broadening) we find the typical linewidth for all observed transitions to be 9(3)\,\,MHz.  The absolute PA laser frequency was measured on a HighFinesse WS-7 wavemeter. The PA laser was stabilized using the wavemeter reading with a slow software feedback (${\sim}\,10\,\,$Hz) to the Ti:Sapphire laser control interface. We have tested the wavemeter performance over a broad range of wavelengths using laser cooling reference light from Li (671\,\,nm), Na (589\,\,nm), and Rb (780\,\,nm), and found our absolute frequency uncertainty to be ${\pm}\,10$\,\,MHz.

\begin{figure}
\includegraphics[width=0.5\textwidth]{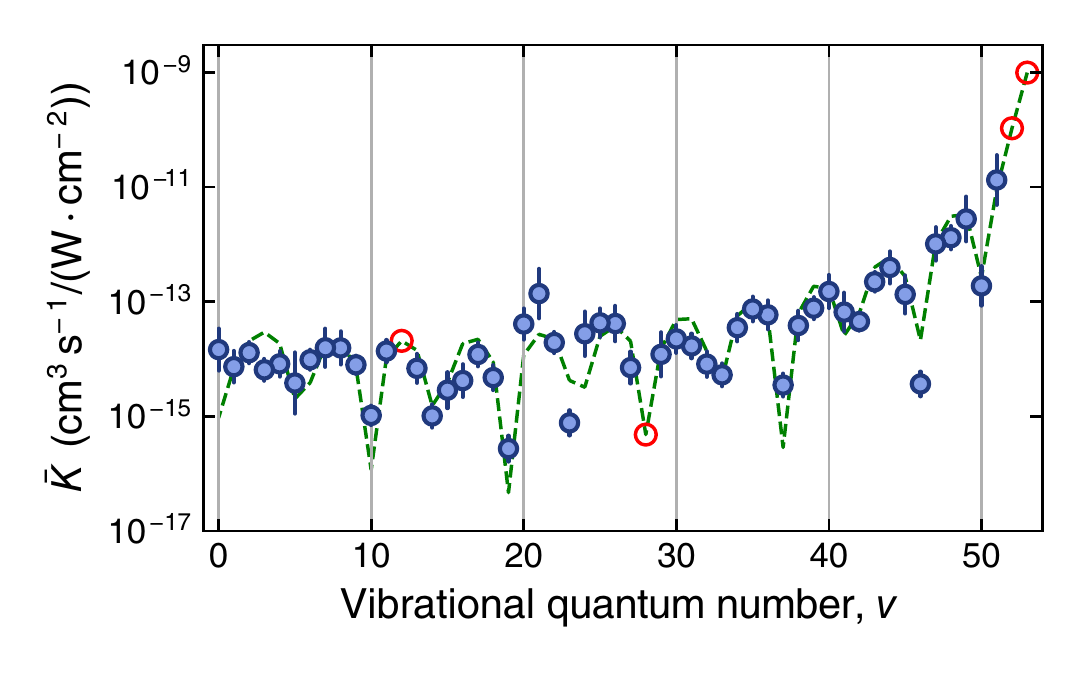}
\caption{PA loss rate coefficients to the $c^3\Sigma^+$ potential. Loss coefficients are shown (blue circles) for the strongest line in each vibrational state multiplet. Red circles indicate lines which were not found in the spectroscopic search. The error bars arise from the fitting uncertainty between the two models (eq.\eqref{eq:twobodyloss}, \eqref{eq:linloss}) and the in-trap atomic density variation. The loss data is compared to Franck-Condon factors (green dashed line) calculated between a low temperature scattering wavefunction in the ground $a^3\Sigma^+$ potential and the excited $c^3\Sigma^+$ bound wavefunctions. The only fit parameter in the \textit{ab initio} theory is the overall proportionality factor between the Franck-Condon factors and the loss rates, $\bar K_v$.}\label{fig:FC}
\end{figure}

Using this technique we have found 50 NaLi PA lines and their substructure (Fig. \ref{fig:allspec}).  We infer the vibrational quantum number, $v$, of these lines from \textit{ab initio} theory calculations of the $c^3\Sigma^+$ potential \cite{OlivierDulieu}, which match the vibrational state binding energies to within 2\,\% for $v\,{\leq}\,49$. The rotational quantum number of these states is $N=1$, which is determined by the selection rule $\Delta N = 1$ for $\Sigma \leftrightarrow \Sigma$ molecular transitions. The use of a single hyperfine state of Li provides a convenient feature in suppressing the PA of Li: Its fermionic character forbids $s$-wave collisions between Li atoms, and hence we do not observe any Li$_2$ PA background compared to other experiments performed in a magneto-optical trap \cite{liyb_ricky}. We observed PA loss features for nearly all vibrational bound states in the $c^3\Sigma^+$ potential, with four exceptions: $v\,{=}\,52, 53$ were too close to dissociation, and thus the ultracold gas would rapidly heat due to off-resonant Rayleigh scattering involving the Li D1 line, $v\,{=}\,28$ which we expect was difficult to detect due to a small Franck-Condon factor (see Fig.\,\ref{fig:FC}), and $v\,{=}\,12$ which is coincident with a water vapor absorption line preventing the Ti:Sapphire laser from lasing.

The creation of ro-vibrational ground state $a^3\Sigma^+$ NaLi molecules relied on the $v\,{=}\,11$ level as an intermediate state in a two-photon transfer, for which identification of state fine structure was necessary.  The fine splitting in the $v\,{=}\,11$ state was studied by observing line strengths under different polarizations of PA light, and considering angular momentum selection rules. Based on the initial collisional state $|J=1, m_J=1\rangle$, the $J=1,2,0$ structure (Fig.\,\ref{fig:sampleline}a) was established by observing PA for different polarizations: $\sigma^+$ light can excite only $J\,{=}\,2$, $\pi$ light cannot excite $J\,{=}\,0$ (producing $\pi$ polarization required using an orthogonal PA beam direction, transverse to the long axis of the trap). The ordering of $J$ states is due to the electronic spin-spin and spin-rotation interactions given by the Hamiltonian \cite{bobfield_book} 
\begin{equation}
H_\text{exc}=A\left(3(\vec S_1 \cdot \hat r)(\vec S_2 \cdot \hat r) - \vec S_1 \cdot \vec S_2\right)+B \vec S \cdot \vec N
\end{equation}
where $\vec S_{1,2}$ are the individual electron spins, $\vec S=\vec S_1+\vec S_2$, $\hat r$ is the unit vector between the two electrons, and $A, B$ are coupling constants (the $A$ constant is an empirical value and contains the $r^{-3}$ dependence of the spin-spin interaction). The observed line structure in $v\,{=}\,11$ is described by $A\,{=}\,182.1(1)$\,\,MHz, and $B\,{=}\,16.2(1)$\,\,MHz. This Hamiltonian and line structure is similar to that of triplet O$_2$, but with opposite energy ordering of $J$ states. This is explained by the different relative orientations of the two electron spins in the $\sigma$ bond of NaLi compared to the $\pi$ bond in O$_2$, resulting in opposite signs in the spin-spin coupling constants \cite{kramers_o2spec,schlapp_o2spec}. The value of the $A$ constant is consistent with a purely magnetic interaction between two electron spins separated by the bond length of NaLi.

Alongside vibrational line positions, we have measured the line strengths on resonance with the strongest observed line in each vibrational multiplet. The PA laser intensity was varied with a fixed exposure time, and the atom number loss as a function of pulse area was used to extract a two-body loss coefficient (Fig.\,\ref{fig:sampleline}b). By increasing the PA laser intensity from zero (rather than exposure time), we ensure that the initial PA loss is at low laser intensities and avoid the effect of PA saturation, simplifying the data analysis \cite{bohn_julienne_PAsat}. With the PA laser on resonance, the free atom loss is governed by the two-body rate equations:
\begin{equation}\label{rateEq}
\dot n_\text{Na}=\dot n_\text{Li}=-\bar K_v I n_\text{Na}n_\text{Li}\\
\end{equation}
where $n_\text{Na},n_\text{Li}$ are the average free atom densities, $I$ is the PA laser intensity, and $\bar K_v$ is the intensity normalized PA rate coefficient. The total density $n\equiv n_\text{Na}+n_\text{Li}$ has the solution:
\begin{align*}
\displaystyle\dfrac{n(t)}{n(0)} & =\left(\displaystyle\dfrac{1-R}{1+R}\right) \left(\displaystyle\dfrac{1+R e^{\delta I t}}{1-Re^{\delta I t}}\right) \numberthis \label{eq:twobodyloss} \\ 
 & \approx 1-\dfrac{2n_\text{Na}n_\text{Li}}{n_\text{Na}+n_\text{Li}}\Big|_{t=0}\bar K_vIt, \quad \delta It \ll 1 \numberthis \label{eq:linloss} 
\end{align*}
where $It$ is the PA pulse area, $R=n_\text{Na}/n_\text{Li}|_{t=0}$ is the density ratio, $\delta \equiv (n_\text{Na}-n_\text{Li})|_{t=0} \bar K_v$, and eq.\,\eqref{eq:linloss} is a simple linearized loss model valid when the PA loss is small. We explicitly measure the free atom number of each species, and a conversion to average density can be made using an effective volume where the spatial distributions of both species overlap, i.e. $V_\text{eff}=N_\text{Na}N_\text{Li}/\int n_\text{Li} n_\text{Na} dV$. Fig. \ref{fig:sampleline}b shows sample loss data for the $v\,{=}\,11$ line that is fit to the full two-body loss curve (eq.\,\eqref{eq:twobodyloss}) and the linear model (eq.\,\eqref{eq:linloss}) for $n(t)/n(0) \,{>}\, 0.7$, where the dynamic density dependence of the PA can be ignored and a linear model is justified. We performed loss measurements for each vibrational level $v$, and the PA rate coefficients $\bar K_v$ are shown in Fig.\,\ref{fig:FC}. The loss rates are proportional to the free-to-bound Franck-Condon factor, and the line strength data is compared to calculated Franck-Condon factors in \textit{ab initio} potentials showing good agreement. We did not observe significant AC Stark shifts in the PA spectra, and we estimate the average shift over all vibrational levels to be ${<}\,0.1\,\text{MHz}/\text{W}\,{\cdot}\,\text{cm}^{-2}$. 

\section{Near-dissociation expansions and potential fit}

We approach fitting the PA line positions in two ways: A near-dissociation expansion valid only for highly excited vibrational states and an $X$-representation of the entire potential which is seeded with \textit{ab initio} calculations. In the initial stages of PA spectroscopy, we found the LeRoy-Bernstein near-dissociation expansion to be of great utility in providing a simple model to predict vibrational state positions \cite{leroy_bernstein}. No spectroscopic data was available on the excited triplet states of NaLi prior to this work, thus we began the spectroscopic search near dissociation where the vibrational energy spacing is small (${\sim}\,50$\,GHz) to minimize the search range necessary to find an initial PA signal. After first observing PA to $v=51$, the LeRoy-Bernstein expansion and higher-order corrections were used to determine positions of deeper bound vibrational states \cite{comparat}. In the case of NaLi, the excited tail is given by a $C_6$ interaction and the LeRoy-Bernstein formula is:
\begin{equation}
E_\text{b}=-\left[\xi\dfrac{(v_D-v)}{C_6^{1/6}}\right]^3,\, \xi\equiv 2\hbar \sqrt{\dfrac{2\pi}{\mu}}\dfrac{\Gamma(7/6)}{\Gamma(2/3)}
\end{equation}
where $E_\text{b}$ is the binding energy, $\mu$ is the reduced mass, $C_6$ is the long range van der Waals coefficient, $v_D$ is a free fitting parameter which can be interpreted as the fractional vibrational quantum number at dissociation, and $\Gamma$ is the Gamma function. The results of such an expansion are accurate to ${<}\,10\%$ so long as the internuclear separation is larger than the LeRoy radius, $R_\text{LR} \,{=}\, 2\left[\langle r_\text{Na}^2\rangle^{1/2}+\langle r_\text{Li}^2\rangle^{1/2}\right]$, where $r_\text{Na},r_\text{Li}$ are the atomic radii of the constituent atoms \cite{leroy_radius}. Using hydrogen-like wavefunctions with a quantum defect model, we find $R_\text{LR}\,{=}\,13\,\,\si{\angstrom}$ for NaLi, which corresponds to a fitting region of $v\,{\geq}\,48$. In this range, the LeRoy-Bernstein analysis gives $C_6=8.13 \times 10^7\,{\pm}\,10\%\,\,\text{cm}^{-1}\si{\angstrom}^{6}$, which is consistent with theoretical predictions \cite{bussery_c6_excited,sadeghpour_c6_excited}. We have not considered the effect of molecular rotation in this analysis, however the $N\,{=}\,0\leftrightarrow1$ rotational splittings are calculated to be ${<}\,2\%$ of the binding energy for $v\,{=}\,48\,{-}\,51$ and do not significantly affect the measured $C_6$. \
\begin{table}[!htb]
	\caption{NaLi $c^3\Sigma^+$ $X$-representation potential fit parameters. Also given are the functional forms of the $X$-representation, which is segmented into an inner, middle, and outer region (for details, see \cite{nali_Xsigma_2}). Parameters marked with $(*)$ were varied during the fitting procedure.}
	\begin{ruledtabular}\begin{tabular}{lc}\label{table:fitparams}
			Parameter & Value \\
			\hline\hline
			\multicolumn{2}{c}{$r<R_i=1.76322\,\,\si{\angstrom}$} \\ 
			\multicolumn{2}{c}{$V(r)=A+\dfrac{B}{r^{N_s}}$} \\
			$A\ (\text{cm}^{-1})\ (*)$ & $-5.35670 \times 10^{4}$ \\
			$B\ (\text{cm}^{-1}\si{\angstrom}^{N_s})\ (*)$ & $1.25377 \times 10^{5}$  \\
			$N_s$ & $1$  \\
			\hline
			\multicolumn{2}{c}{$R_i \leq r \leq R_o=11.4620\,\,\si{\angstrom}$} \\ 
			\multicolumn{2}{c}{$V(X)=\displaystyle\sum^{18}_{i=0} a_i X^i,\,\, X=\dfrac{r-R_m}{r+bR_m}$}\\
			$R_m\ (\si{\angstrom}) (*)$ & $3.44547$ \\
			$b\ (*)$ & $4.21443\times 10^{-1}$\\
			$a_0\ (\text{cm}^{-1})\ (*)$ & $-4.77774 \times 10^{3}$  \\
			$a_1\ (\text{cm}^{-1})\ (*)$ & $1.34741 \times 10^{-4}$  \\
			$a_2\ (\text{cm}^{-1})\ (*)$ & $5.61284 \times 10^{4}$  \\
			$a_3\ (\text{cm}^{-1})\ (*)$ & $-7.42856 \times 10^{4}$  \\
			$a_4\ (\text{cm}^{-1})\ (*)$ & $-1.34517 \times 10^{4}$  \\
			$a_5\ (\text{cm}^{-1})\ (*)$ & $9.77977 \times 10^{3}$  \\
			$a_6\ (\text{cm}^{-1})\ (*)$ & $-9.25786 \times 10^{4}$  \\
			$a_7\ (\text{cm}^{-1})\ (*)$ & $1.80252 \times 10^{5}$  \\
			$a_8\ (\text{cm}^{-1})\ (*)$ & $2.10411 \times 10^{5}$  \\
			$a_9\ (\text{cm}^{-1})\ (*)$ & $-1.61060 \times 10^{5}$  \\
			$a_{10}\ (\text{cm}^{-1})\ (*)$ & $-1.04454 \times 10^{5}$  \\
			$a_{11}\ (\text{cm}^{-1})\ (*)$ & $2.07198 \times 10^{4}$  \\
			$a_{12}\ (\text{cm}^{-1})\ (*)$ & $-1.04511 \times 10^{5}$  \\
			$a_{13}\ (\text{cm}^{-1})\ (*)$ & $-2.96635 \times 10^{5}$  \\
			$a_{14}\ (\text{cm}^{-1})\ (*)$ & $1.95118 \times 10^{5}$  \\
			$a_{15}\ (\text{cm}^{-1})\ (*)$ & $-6.84050 \times 10^{5}$  \\
			$a_{16}\ (\text{cm}^{-1})\ (*)$ & $-1.97954 \times 10^{5}$  \\
			$a_{17}\ (\text{cm}^{-1})\ (*)$ & $1.41806 \times 10^{6}$  \\
			$a_{18}\ (\text{cm}^{-1})\ (*)$ & $6.28720 \times 10^{5}$  \\

			\hline
			\multicolumn{2}{c}{$R_o < r$} \\
			\multicolumn{2}{c}{ $V(r)=-\dfrac{C_6}{r^6}-\dfrac{C_8}{r^8}-\dfrac{C_{10}}{r^{10}}-E_\text{ex},\,\, E_\text{ex}=B_\text{ex}r^\alpha e^{-\beta r}$}\\
			$C_6\ (\text{cm}^{-1}\si{\angstrom}^{6})\ (*)$ & $7.81853 \times 10^{7}$  \\
			$C_8\ (\text{cm}^{-1}\si{\angstrom}^{8})\ (*)$ & $8.97063 \times 10^{8}$  \\
			$C_{10}\ (\text{cm}^{-1}\si{\angstrom}^{10})\ (*)$ & $9.62572 \times 10^{9}$  \\ 
			$B_\text{ex}\ (\text{cm}^{-1}\si{\angstrom}^{-\alpha})$ & $3.44171 \times 10^{3}$  \\
			$\alpha$ & $5.22956$  \\
			$\beta\ (\si{\angstrom}^{-1})$ & $2.12190$ 		
	\end{tabular}\end{ruledtabular}
\end{table} 

Full potential fitting was performed using the $X$-representation model, which is a piece-wise parametrization with a combination of phenomenological parameters and physical constants, such as oscillator frequency, potential depth, and dispersion coefficients \cite{nali_Xsigma_2}. The initial $X$-representation fit was acquired by minimizing the root-mean-square (RMS) deviation for each point of the \textit{ab initio} potential \cite{OlivierDulieu}. The details of the fitting procedure can be found in \cite{ourtwophoton}. Then, this $X$-representation was further optimized through simulated annealing using the RMS error of the vibrational binding energies as the objective function.

The results of the simulated annealing are reported in Table\, \ref{table:fitparams}\,; the free parameters are marked with an asterisk. The vibrational binding energies calculated from this improved $X$-representation potential give an RMS error from the measured values of 7\,\,GHz, a factor of 30 improvement over the \textit{ab initio} potential. The fitting process shifted the van der Waals coefficients $C_6$ and $C_8$ by $-4 \%$ from their theoretical values \cite{sadeghpour_c6_excited}. These coefficients only have strong impact on the energies of states that have substantial perturbation from spin-orbit coupling to the $^1 \Pi$ potential. Since the $\Pi$ dispersion coefficients are predicted to be much smaller \cite{sadeghpour_c6_excited}, it is not surprising that we find values slightly below the theoretical value for the $\Sigma$ potentials' coefficients. There was no theoretical prediction of the higher-order dispersion coefficient, $C_{10}$.
The exchange energy terms ($B_\text{ex},  \alpha,$ and $\beta$, see Table\, \ref{table:fitparams}) were estimated from the values reported for the $a^3\Sigma^+$ potential \cite{nali_Xsigma_2, russiantheory} and kept as constants for the fitting procedure since their contribution to the potential energy was less significant than those of the dispersion terms. Finally, $B_{ex}$ was slightly modified in order to improve the continuity/differentiability of the potential.

\section{Conclusion and Outlook}

We have performed heteronuclear PA spectroscopy from an initial ultracold mixture of Na and Li, mapping out 50 vibrational states in the excited $c^3\Sigma^+$ potential. By measuring the free atom loss rate from PA, we determined the free-to-bound line strengths, showing good agreement with \textit{ab initio} theory. This work was used in formation of $a^3\Sigma^+$, $v\,{=}\,0$, $N\,{=}\,0$ NaLi ro-vibrational ground state molecules, where a Feshbach resonance was used to access a weakly bound molecular state followed by two-photon transfer to the ground state via the $c^3\Sigma^+$, $v\,{=}\,11$, $N\,{=}\,1$ excited state. In similar efforts to produce high phase-space density samples of heteronuclear dipolar molecules, the magneto-association suffers from low molecule formation efficiency \cite{feshmol_ketterle_2012,krb_feshbach,nak_feshbach,rbcs_feshbach,rbcs_feshbach2,narb_feshbach}, unless an optical lattice is used to mitigate inelastic loss \cite{krb_lattice_2012}. It has been recently demonstrated that a coherent, all-optical transfer to a molecular state is possible \cite{free2bound_stirap}. Using the spectroscopic data reported in this work could allow for a similar all-optical molecular formation scheme that would bypass the use of Feshbach resonances, which are particularly narrow in the NaLi system. 

\section*{Conflicts of interest}
There are no conflicts to declare. 

\begin{acknowledgments}
We would like to thank Li Jing and Yijun Jiang for experimental assistance. We acknowledge support from the NSF through the Center for Ultracold Atoms and award 1506369, from the ONR DURIP grant N000141613141, and from MURIs on Ultracold Molecules (AFOSR grant FA9550-09-1-0588) and on Quantized Chemical Reactions of Ultracold Molecules (ARO grant W911NF-12-1-0476). H.S. and J.J.P. acknowledge additional support from the Samsung Scholarship. 
\end{acknowledgments}

\bibliographystyle{apsrev4-1}
%

\end{document}